\documentclass{iopconfser}
\usepackage{graphicx}%
\usepackage{multirow}%
\usepackage{amsmath,amssymb,amsfonts}%
\usepackage{amsthm}%
\usepackage{mathrsfs}%
\usepackage[title]{appendix}%
\usepackage{xcolor}%
\usepackage{textcomp}%
\usepackage{manyfoot}%
\usepackage{booktabs}%
\usepackage{algorithm}%
\usepackage{algorithmicx}%
\usepackage{algpseudocode}%
\usepackage{listings}%
\usepackage{bbold}
\usepackage{amsmath}
\usepackage{amsthm}
\usepackage{amssymb}
\usepackage{amsfonts}
\usepackage{tensor}
\usepackage{braket}
\usepackage{slashed}

\theoremstyle{definition}

\newcommand{\vect}{\textbf}

\newcommand{\vd}{\boldsymbol{\partial}}

\begin{document}

\title{The Dirac equation and the Quantum Potential}

\author{Calum Robson$^{1}$}

\affil{$^1$Department of Mathematics, LSE, London, UK}

\email{c.j.robson@lse.ac.uk}

\begin{abstract}
One key theme of Basil Hiley's work was the development of David Bohm's approach to Quantum Mechanics; in particular the concept of the quantum potential. Another theme was the importance of Clifford Algebras in fundamental physics.   In this paper I will combine these approaches by looking at how the quantum potential can be extended to the Dirac equation. I will begin by discussing the geometry of the Dirac equation, and how this is made clearer by the use of Clifford algebras . Next, I will rewrite the Cl(2) Dirac wavefunction in Polar form, and show that new behaviour arises due to topological nonlocality. Finally, I discuss the relationship between the Dirac and Schroedinger equations. 
\end{abstract}

	\section{Introduction}
The Dirac equation is one of the most important equations in Physics, and also one of the most mysterious-- this extends even to its origins, apparently occurring to Dirac in a dream\footnote{Of an ouroboros, no less. For an overview of the full history, see \cite{Schweber1994}}. In this paper, I will begin to discuss the interpretation of the Dirac equation within the Bohm/Hiley approach to Quantum Mechanics, and using mathematics of Geometric Clifford Algebras. Before beginning , I should acknowledge my debt to Basil Hiley, and explain how his ideas are connected to the topic of this paper. \\
I first came across Basil's work when I was a PhD student. I was studying noncommutative Instantons, which are described using a biquaternion algebra, and this lead me to the study of Clifford algebras. At the same time one of the canons at Durham Cathedral suggested I read David Bohm's, `Wholeness and the Implicate Order' \cite{Bohm1980}. I did this, and I was convinced by the account given there of the Implicate and Explicate orders, and also by the argument that algebras were the correct mathematical object for describing fundamental physics\footnote{Because they have an addition operation, allowing for comparison and ratio, and a multiplication operation, which can be used to describe interaction.}. One of Basil's main insights, as I see it, is that Clifford algebras are a powerful language to describe fundamental physics. As I began to learn more about Clifford algebras, after completing my PhD, I kept coming across papers by Basil, and I read a few interviews with him-- but I never suspected that he was not only still active in research, but producing some of his best work. \\
I was introduced to Basil\footnote{ by Mike Wright, of the Archive Trust} shortly after I moved to London at the end of the pandemic, to begin teaching at the LSE. I remember his deep knowledge of mathematics and physics, his patience in explanation, and his enthusiasm for life. His lifelong search for scientific truth is an inspiration, as is his unfailing curiosity and determination. Basil's example confirmed me in my decision to continue with my research, and though the time I knew him was short, his influence on me was immense. \\
I believe that Basil's approach to fundamental physics-- based around Clifford Algebras, and a quantum phase space-- is the correct one, and I hope to continue this research programme. With that in mind, I should explain how I think this talk is relevant to the approach to physics begun by David Bohm, and expanded upon and built up by Basil Hiley. As I see it, this has five main features, from the most abstract to the most concrete. 
\begin{enumerate}
	\item The way forward for physics requires new conceptual ideas, not simply new mathematics or experimental data (though these are also probably necessary).
	\item It is both possible and desirable to find an, `ontological' interpretation of Quantum Mechanics, as opposed to treating it in a positivist way as a theory which predicts experimental results but can say nothing about the underlying reality. 
	\item The main idea that is missing from our current theories is holism; and this is best understood as an Implicate order, which cannot be known in itself but, `unfolds' as different Explicate orders relative to different ways of trying to know it (for example wavelike or particlelike behaviour in response to different experimental setups).
	\item This underlying holism has two main consequences. First, fundamental physics follows a process ontology, represented by an algebra. Second, the world is fundamentally nonlocal, and this is the key message of quantum mechanics as opposed to classical physics.
	\item Finally, the polar decomposition of the wavefunction and the corresponding, `quantum potential' which emerges from the Schroedinger equation are vital, if incomplete, clues toward a deeper theory
\end{enumerate}
How does this paper fit within this perspective? There are two main ways. First of all, it is based on the use of Clifford algebras. It discusses the Dirac operator as the differential operator which is inseparable from Clifford algebras, and which therefore must be well understood in order to make full use of them. A major theme of this paper is that the Dirac equation is fundamentally nonlocal, because its solutions are determined by the (nonlocal) topological features of the space the equation is defined over. This aligns with the focus on nonlocality in the Bohm/ Hiley approach but gives new examples of nonlocality arising from topology.\\
I begin by using the geometric algebra representation of Clifford algebras to interpret the action of the Dirac operator in terms of flux. I begin by looking at free solutions of the form $\vd Z=0$, before looking at the implications of adding a potential, focussing on vector potentials $\vect{A}$ which arise from gauge theories. I discuss how (some) solutions to the Dirac equations in this context are determined by topology, using results from homology theory.\\
Secondly, given the importance of the Schroedinger equation and the quantum potential to the Bohm/ Hiley approach to quantum mechanics, I then move on to discuss the relationship between the Dirac and Schroedinger equations, suggesting avenues for future work. I then conclude the paper by analysing the action of the Dirac equation in $Cl(2)$ on a wavefunction in the polar decomposition, and show that this leads to some interesting physical behaviour. \\
First, though, I will outline the geometric algebra approach, which underpins this paper mathematically, and briefly review some necessary results from algebraic topology. 
\subsection{Geometric Algebras}
Throughout this paper, I will use the language of geometric algebra, which is a method of working with Clifford algebras by assigning geometric meaning to the elements\footnote{Conversely, it is a method of analysing geometry by associating geometric objects to elements of a Clifford algebra}. There are different ways to associate geometry to Clifford algebras\footnote{The other method, popular with computer scientists, involves projective geometries \cite{Lengyel2024}}. I shall here use the original method developed by David Hestenes \cite{Hestenes1987} and expanded upon by various other authors \cite{Gull1993}\cite{Doran2003}\cite{Macdonald2010}. There are competing accounts of the significance of geometric algebra, so I will start by giving my own. \\
We start by analogy with the complex plane. What is the difference between $\mathbb{R}^{2}$ and $\mathbb{C}$? The answer is that one is a vector space and the other is an algebra. In $\mathbb{R}^{2}$, we can add and subtract vectors, and multiply them by scalars-- but we cannot add them. On the other hand, $\mathbb{C}$ has the same addition as $\mathbb{R}^2$ but also has a multiplication operation. If we define
\begin{equation}
	z_{1}=\rho_{1}e^{i\theta_{1}}; \ \  z_{2}=\rho_{2}e^{i\theta_{2}}
\end{equation}
then we have
\begin{equation}
	z_{1}z_{2}=\rho_{1}\rho_{2}e^{i(\theta_{1}+\theta_{2})}
\end{equation}
The lengths of the vectors multiply, and the angles add. We can see from this why $\sqrt{-1}=i$, since we can write $-1$ on the $x-$axis as $e^{i\pi/2}$, and the number that squares to give this is $e^{i\pi/4}$. Geometrically this is the point which is  at radius 1 on the $y$-axis; which is the point $i$ on the complex plane.\\
Geometric algebras arise when we do the same thing for $\mathbb{R}^{p,q}$-- moving from a vector space to an algebra by defining an product on the elements. If we have an orthonormal basis $\{e_{i}\}$, we define the product of different vectors $e_{i}e_{j}$ to be the area spanned by $e_{i}$ and $e_{j}$, the product $e_{i}e_{j}e_{k}$ to be the 3-volume spanned by the three different component basis vector; and so on. The product of all the orthonormal basis vectors $e_{1}e_{2}...e_{n}$ is usually denoted $I$, and called the pseudoscalar, or volume form\footnote{So far, this is simply the exterior algebra. The Clifford algebraic structure comes from introducing the metric, directly below}. \\
If vectors are not distinct, then we have
\begin{equation}\label{Clifforddef}
	e_{i}e_{j}+e_{j}e_{i}=2\delta_{ij}
\end{equation}
Note that this implies that orthogonal vectors anticommute; $e_{i}e_{j}=-e_{j}e_{i}$ if $i\neq j$. We shall work with orthonormal bases in this paper, however in general, for a non- orthonormal basis we have 
\begin{equation}\label{Clifforddef}
	e_{i}e_{j}+e_{j}e_{i}=2g_{ij}
\end{equation}
This is the defining relation of a Clifford algebra \cite{Lounesto2009}, where $g_{ij}$ is the metric. We call the number of basis vectors in an element the \textsl{grade} of that element. So a scalar is grade 0, a vector is grade 1, and area is grade 2, and so on. The highest grade object is the pseudoscalar, which is grade $n$. \\
These definitions extend to a general vector $\vect{v}$ by linearity. Then, given two vectors $\vect{v}$ and $\vect{w}$, we can define an inner product $\vect{v}\cdot\vect{w}$ that returns the scalar part of the product $\vect{v}\vect{w}$, and an exterior product $\vect{v}\wedge\vect{q}$ which returns the grade 2 (or bivector) part. This generalises to higher grade objects (or multivectors). If $A$ is grade $p$ and $B$ is grade $q$, with $p\leq q$ then we can form an inner product\footnote{There are some technical details here -- see \cite{Lengyel2024}} $A\cdot B$ which has grade $q-p$ , and an outer product $A\wedge B$ with grade $p+q$. \\
Geometrically, $A\cdot B$ is spanned by the basis vectors in $B$, but not in $A$, and $A\wedge B$ is spanned by the basis vectors in $A$ or $B$. It is often more useful to have the basis vectors contained in $A$ and $B$. This is the projection of $A$ into $B$, defined by 
\begin{equation}
	P_{B}(A)\equiv(A\cdot B)B^{-1}
\end{equation}
The complement of $A$ in $B$, sometimes called the rejection of $A$ from $B$, is given by
\begin{equation}
	R_{B}(A)\equiv (A\wedge B) B^{-1}
\end{equation}
To evaluate an expression like $AB$ we simply write both multivectors next to each other, and use the identity (\ref{Clifforddef}) to simplify the resulting products of basis vectors. \\
The geometric algebras most useful for our case here are $Cl(2)$ and $Cl(3,1)$. First, $Cl(2)$ is given by 
\begin{equation}
	\{ 1, e_{0}, e_{1}, e_{0}e_{1} \}
\end{equation}
where $e_{0}e_{1}=I$, the pseudoscalar, and $I^{2}=-1$. Therefore the even subalgebra $Spin(2)=\{1, I\}\cong\mathbb{C}$. This is a useful Clifford Algebra to test concepts in, since intuition from complex analysis can often be employed directly. The algebra $Cl(3,1)$ is sometimes called the Spacetime algebra, since it is the geometric algebra generated by Minkowski space. It has a basis given by
\begin{equation}
	\Big\{ 1, \gamma_{a}, \gamma_{0}\gamma_{i}, \gamma_{i}\gamma_{j}, \gamma_{a}\gamma_{b}\gamma_{c}, I \Big\}
\end{equation}
Here $\{a,b,c \}$ run from 0 to 4, whereas $\{i,j,k\}$ refer to the spacelike directions, $\{1,2,3\}$. This algebra contains four vectors $\gamma_{a}$, three timelike rotations (or boosts) $\gamma_{0}\gamma_{i}$; three spacelike rotations $\gamma_{i}\gamma_{j}$; four $3-$volumes $\gamma_{a}\gamma_{b}\gamma_{c}$; and the volume form $I\equiv\gamma_{0}\gamma_{1}\gamma_{2}\gamma_{3}$. We use $\gamma_{a}$ to denote the basis, since the usual gamma matrices in Dirac theory form a representation of this algebra\footnote{by definition, the matrix $\gamma_{5}$ represents  the pseudoscalar $I$}. Indeed, the ability to give a clear geometric interpretation to the gamma matrices is one of the main advantages of using geometric algebras \cite{Gull1993}. 
\\
For any Clifford algebra, we can define a Dirac operator by
\begin{equation}
	\vd\equiv e^{i}\partial_{i}
\end{equation}
Where the reciprocal basis $e^{i}$ satisfies $e^{i}e_{j}=\delta^{i}_{j}$. In this paper we will work with orthonormal bases and so $e^{i}=e_{i}$. To see why we call this the Dirac operator, note that for Minkowski space, and $Cl(3,1)$, we have
\begin{equation}
	\vd=\gamma_{0}\partial_{0}+\gamma_{1}\partial_{1}+\gamma_{2}\partial_{2}+\gamma_{3}\partial_{3}
\end{equation}
Once we have the Dirac operator $\vd$, we can form two derivatives of a multivector $A$ of grade $k$. There is the interior derivative $\vd\cdot A$, and the exterior derivative $\vd\wedge A$. I shall discuss the geometric interpretation of these operators below. \\
An important property of Dirac operators is that their solutions are also solutions of Laplace's equation, and hence they are harmonic functions. To see this, note that for an orthonormal \footnote{for a non- orthonormal basis the usual modifications to this formula apply-- see \cite{Macdonald2010}) } basis $\vd\cdot\vd=\sum_{i}\partial^{2}_{i}\equiv\Delta^{2}$. Therefore any solution to $\vd Z=0$ will also be a harmonic function, satisfying $\Delta^{2} Z=0$. 
\\
This is an extension of the link between holomorphic functions in complex analysis and harmonic functions. In fact, we can generalise complex analysis to the Clifford algebra case\cite{Delanghe1992}. The resulting area of mathematics is known as Clifford analysis. We call solutions to $\vd Z=0$ \textsl{Monogenic Functions}. These are a generalisation of holomorphic functions. To see this, we can work in $Cl(2)$. Here, the Dirac operator is $\vd=e_{1}\partial_{1}+e_{2}\partial_{2}$. Let this act on a function $Z=u+Uv\in Spin(2)\cong\mathbb{C}$. Then
\begin{align}\nonumber
	\vd\big(u(\tilde{z}))+Iv(\tilde{z})\big)=e_{1}\frac{\partial u}{\partial x}+e_{2}\frac{\partial u}{\partial y}+e_{1} I\frac{\partial v}{\partial x}+e_{2}I\frac{\partial v}{\partial y}\\
	=e_{1}\Big(\frac{\partial u}{\partial x}-\frac{\partial v}{\partial y}\Big)+e_{2}\Big(\frac{\partial u}{\partial y}+\frac{\partial{v}}{\partial x}\Big)
\end{align}
We can recognise the brackets in the last line from the Cauchy-Riemann equations. Then the function will be monogenic iff it is holomorphic. \\
Another important result that generalises is the Cauchy theorem, which becomes, for a general Clifford algebra
\begin{equation}
	f(y)=\frac{1}{I S_{n}}\oint_{\partial V}\frac{x-y}{\lvert x-y\rvert^{n}}f(x) 
\end{equation}
We can see this as defining a Green's function for the Dirac operator
\begin{equation}
	G(x-y)=\frac{1}{S_{n}}\frac{x-y}{\lvert x-y\rvert^{n}}; \ \vd G =\delta(x-y)
\end{equation}
To me, this suggests that many of the times we use complex analysis in theoretical physics-- for example, contour integrals in QFTs-- we might actually be better off using Clifford analysis and the spacetime algebra. However this remains unstudied. 
\subsection{Some Algebraic Topology: Homology and Homotopy }
A discussion of algebraic topology in a paper like this must be either very long or very short. I shall aim for the second. For more details, see e.g. \cite{Hatcher2001}\cite{Bott1982}.   \\
Topology is the study of features of shapes or manifolds which cannot be changed under continuous transformations. Famously, a doughnut and a coffee mug are topologically identical, since they both have one hole (we can imagine turning a soft clay doughnut into a clay mug-- the central hole of the doughnut becomes the handle of the mug). \\
In fact, the features of shapes which cannot be changed by continuous transformation tend to be either boundaries or holes\footnote{And any hole must have a boundary, so we can really just consider boundaries}. Boundaries and holes can be of any dimension-- a circle is a 1d boundary, and the interior of a hollow sphere $(S^{2})$ is a 3d hole\footnote{The 2d surface of the sphere is then the boundary of this 3d hole}.\\
Algebraic topology makes analysing these structures easier by assigning groups to them in a consistent way\footnote{Using Category Theory}. There are two kinds of  groups we will focus on here, with confusingly similar names. These are homology groups, and homotopy groups. \\
The homology groups of a manifold $\mathcal{M}$, denoted $H_{k}(\mathcal{M})$, can be thought of as counting the boundary components of $\mathcal{M}$ of dimension $k$. In particular, the rank of $H_{k}$, called the Betti number, and denoted $\beta_{k}$, is equal to the number of boundary components of $\mathcal{M}$ of dimension $k$. \\
There are many ways to define and determine the homology groups. For our purposes, the key result comes from an area of mathematics called Hodge theory\cite{Schwarz1991}\cite{Roberts2022}. For compact, euclidean, manifolds, the homology groups $\mathcal{H}_{k}$ are the spaces of harmonic $k$-forms; that is, differential forms $\chi$ of grade $k$ satisfying $\Delta\chi=0$. The importance of this for our purposes arises from the fact that harmonic forms of order $k$ correspond to grade $k$ Clifford algebra functions, and from the fact that $\vd\cdot\vd= \Delta$. This means that harmonic $k$-forms are in one-to-one correspondence with grade $k$ solutions to $\vd Z_{k}=0$. Hence, for a compact, euclidean, manifold, the homology groups determine the kernels of the Dirac operator (and vice versa)\footnote{Note, however, that this has not yet been proven for or extended to Lorentzian manifolds}. \\
Another sort of topological group is the homotopy group. These are defined as $\pi_{k}(\mathcal{M})=Hom(S^{k}, \mathcal{M})$; the space of all maps from $S^{k}$ to $\mathcal{M}$. Since most boundaries are topologically equivalent to some $S^{m}$, and since any map $S^{k}\rightarrow\mathcal{M}$ restricts to a map on each boundary component\footnote{The link to boundary components means that there is a relation between homology and homotopy, given by the Hurewicz theorem \cite{Bott1982}.} , we are often interested in  $S^{k}\rightarrow S^{m}$, which lie in the group $\pi_{k}(S^{m})$. \\
In general, these groups are very hard to calculate, but in this paper I will focus mostly on manifolds with a single boundary topologically equivalent to $S^{1}$. Then $\pi_{k}(\mathcal{M})=\pi_{k}(S^{1})$. In particular $\pi_{1}(S^{1})=\mathbb{Z}$, with the integers in $\mathbb{Z}$ corresponding to the winding number of the map-- how many times it wraps around the boundary $S^{1}$. 
\section{Flux and the Dirac Equation}
The first insight from the Clifford Algebra perspective is that the Dirac equation does not only act on Spinors. Whilst it certainly can do this (and needs to if we want to reproduce the original Dirac equation), it can act on any geometric object $G$ , simply by calculating $\vd G$. In particular, we can write Maxwell's equations in terms of the Dirac operator as
\begin{equation}
	\vd F=J
\end{equation}
For a bivector field strength $F$, and a current $J$. To see a detailed discussion of how the Dirac equation acts on Spinors in the Clifford algebra context, see the work by Callaghan and Hiley \cite{HileyCallaghan2010}\cite{Hiley2011}. For another approach see Hestenes \cite{Hestenes2003}\cite{Hestenes2010}. For many more examples of Dirac Operators in physics\footnote{I have also written a paper discussing some features of the Dirac equation in the Maxwell context-  see \cite{Robson2025}} see the book \cite{Doran2003} . \\
By definition a differential operator $D$ defines some kind of change. The most straightforward equation we can write, $Df=0$, tells us that the object $f$ does not change in the manner represented by $D$. If we now add a potential $V$ so that $Df=V$, this tells us that the change in $f$ is described by the function $V$. In the case of the Dirac equation, we have a vector- valued differential operator $\vd$, which describes a change with both magnitude and direction. Therefore it defines a flux. \\
We can split the Dirac equation $\vd G$ into two parts, $\vd\cdot G$ (which lowers the grade by 1), and $\vd\wedge G$ (which raises the grade by 1). If we were acting on a vector in $\mathbb{R}^{3}$ these would correspond to the divergence and the curl\footnote{More accurately, $ \vd\wedge\vect{v}=(\vd\times\vect{v})I$, the geometric dual of the cross product}. More generally, $\vd\wedge G$ is determined by the parts of $\vd G$ which are orthogonal to $G$, and $\vd\cdot G$ is determined by the parts of $\vd G$ which are parallel to (or lie within) $G$. We can make this clearer by forming the rejection $(\vd\wedge\vect{A})A^{-1}$, which will always be orthogonal to $G$, and the projection $(\vd\cdot A)A^{-1}$, which is will give a sum over the directions $e_{i}$ that lie in $G$. \\ 
Therefore $\vd\wedge A$ is a measure of external flux, and $\vd\cdot A$ determines the flux within $A$, or the internal flux. To illustrate the principle,suppose we have two functions $F$ and $G$ in $\mathbb{R}^{3}$, with orthonormal basis $\{ e_{1}, e_{2}, e_{3} \}$ defined by 
\begin{align}\label{eqn:exampledef} \nonumber 
	F(x_{1}, x_{2})= f(x_{1}, x_{2})\ e_{1}\wedge e_{2} \\ \nonumber
	G(x_{1}, x_{3})= g(x_{1}, x_{3})\ e_{1}\wedge e_{2}
\end{align}
Then
\begin{align}\nonumber
	\vd\cdot F &=f_{,1}e_{2}-f_{,2}e_{1}\\ \nonumber
	(\vd\cdot F)F^{-1}&= \frac{1}{f}\big(f_{,1}e_{1}+f_{,2}e_{2}\big)=\vd\big(\text{ln}(f)\big)\\
	\vd\wedge F &= \big(\vd\wedge F\big)F^{-1}=0
\end{align}
and
\begin{align}\nonumber
	\vd\cdot G &= g_{,1}e_{2} \\ \nonumber
	\big(\vd\cdot G\big)G^{-1}&=\frac{g_{,1}}{g}e_{1}\\ \nonumber
	\vd\wedge G& =g_{,3}\ e_{1}\wedge e_{2}\wedge e_{3}\\
	\big(\vd\wedge G\big) G^{-1}&=g_{,3}e_{3}
\end{align}
So we see that both $F$ and $G$ can  possess internal flux, but only $G$ can have non-zero external flux. Of course, imposing $\vd F=\vd G =0$ forces all these fluxes to vanish. \\
For a compact, Euclidean, manifold $\mathcal{M}$, Hodge Theory shows that the space of solutions to the Dirac equation $\vd Z_{k}=0$ on $\mathcal{M}$ (where $k$ denotes the grade of the multivector function $Z_{k}$) is equal to the k$^{th}$ homotopy group of $\mathcal{M}$, and therefore the dimension of this solution space is equal to the k$^{th}$ Betti number \cite{Roberts2022}. \\
A topologically trivial space, with all homology groups zero,  only allows for a constant uniform flux, as $\vd F=0 \iff F=\ \text{const}$. This is equivalent to a zero field under a suitable rescaling. As explained in the introduction, none- zero homology groups, and so non- constant solutions to the Dirac equation, correspond to topologically non-trivial spaces- i.e. those with boundaries or punctures \footnote{which can be thought of as defining a point-like boundary}. \\
The flux picture of the Dirac equation gives us a way of understanding this. If we want a non- constant flux, it must have sources and sinks. These sources and sinks will be boundary components (through which flux can enter or leave). It therefore makes physical sense that the solution spaces of the Dirac equation correspond to the homotopy groups of the space. \\
I should stress that this physical picture is only fully available in spaces where Hodge theory applies-- that is, compact, euclidean spaces. In particular it is unclear how to rigorously define the topology of Lorentzian spaces; and therefore spacetimes \cite{Papadopoulos2021}. I am, however, hopeful that the use of Clifford algebra methods might throw light on this in the future. 
\subsection{Gauge Fields and Boundaries}
So far we have considered the equation $\vd Z=0$. What happens if we add in a potential? If the equation $\vd Z =0$ states that the overall flux is zero, then as discussed above, $\vd Z = V$ states that the flux is described by the multivector function $V$. It should be noted that in this case, we lose the topological machinery described in the previous section, which only applies when $\vd Z =0$. \\
A special case of a potential which is very important in physics is the vector potential arising from a gauge field, which appears in the Dirac equation as
\begin{equation}\label{eqn:DiracGauge}
	\big( \vd + \vect{A}\big)Z =0
\end{equation}
Rearranging to $\vd Z =\vect{A} Z$, we can split this into the inner and outer products as
\begin{align}\nonumber
	\vd\cdot G=\vect{A}\cdot G \\ 
	\vd\wedge G =\vect{A}\wedge G
\end{align}
We can also calculate
\begin{align} \nonumber 
	\big(\vd\cdot G\big) G^{-1}=\big(\vect{A}\cdot G\big)G^{-1}\equiv  P_{G}(\vect{A})\\	\big(\vd\wedge G\big)G^{-1}=\big(\vect{A}\wedge G \big) G^{-1}\equiv R_{G}(\vect{A})
\end{align}
Where $P_{G}(\vect{A})$ is the orthogonal projection of $\vect{A}$ into $G$, and $R_{G}(\vect(A)$, the rejection of $\vect{A}$ in $G$, is the orthogonal complement of $P_{G}(\vect{A})$ within $\vect{A}$ (i.e. the bit of $\vect{A}$ which is orthogonal to $G$). Using our function $G$ from our previous example, equation (\ref{eqn:exampledef}), let $\vect{A}\equiv A^{1}e_{1}+A^{2}e_{2}+A^{3}e_{3}$. Then we have
\begin{align}\nonumber
	\vd \cdot G&= \vect{A}\cdot G\iff g_{,1}e_{2}=\big(A^{1}e_{2}-A^{2}e_{1}\big)g\\
	\vd\wedge G&=\vect{A}\wedge G\iff g_{,3}\ e_{1}\wedge e_{2}\wedge e_{3}=A^{3} \ g \ e_{1}\wedge e_{2}\wedge e_{3}
\end{align}
or equivalently
\begin{align}\nonumber 
	\big(\vd \cdot G\big)G^{-1}&=  \big(\vect{A}\cdot G\big)G^{-1}\iff \frac{g_{,1}}{g}e_{1}=A^{1}e_{1}+A^{2}e_{2}\\
	\big(\vd\wedge G\big)G^{-1}&=\big(\vect{A}\wedge G\big)G^{1}\iff \frac{g_{,3}}{g}e_{3}=A^{3}\ e_{3}
\end{align}
This is solved by setting $A_{2}=0$, and $g(x_{1}, x_{3})=exp(A^{1}x_{1}+A^{3}x_{3})$\\
In general, then, the gauge field $\vect{A}$ defines a flow of flux. Interestingly, with this interpretation of $\vect{A}$, the Lorentz Gauge, where $\vd\cdot\vect{A}=\partial_{\mu}A^{\mu}=0$, is a conservation equation for the flux current defined by $\vect{A}$. It is puzzling but worth further investigation that the conservation of flux can be gauge- dependent in this way. \\
We get similar gauge dependency on flux if we begin with a a solution $\vd Z=0$ and then gauge- transform to $	\big( \vd + \vect{a}\big)Z =0$, where $\vect{a}$ is pure gauge (i.e. gauge equivalent to zero). Then a solution with zero flow of flux-- including on the boundaries-- becomes, via a gauge transformation, a solution with non- zero flux\footnote{This reminds me of the Unruh effect \cite{Earman2011} in semiclassical gravity, in which a gauge transformation to an accelerated reference frame leads to a thermal bath of particles appearing from the vacuum. I am unsure, however, if there is a formal link between these cases.} determined by $\vect{a}$. \\ 
In the previous section we saw that solutions to the Dirac equation $\vd Z=0$ are determined by the topology of the underlying space. Does such a topological link exist for solutions with a potential?  This is something I am still investigating. The answer seems to lie with the Atiyah-Singer index theorems, which link the kernels and cokernels of differential operators on a manifold to topological invariants of that manifold \cite{Nakahara}\cite{Berline1992}. I am still unsure, however,  how these theorems apply in the Geometric algebra case, though I am actively working on understanding this. That said, there is a well studied sub-category of potentials where we definitely have a topological dependence \cite{Manton2004}.  \\
These are solutions for which the gauge potential $\vect{A}$ is pure gauge on the boundary.  In these cases we can treat $\vect{A}$ on the boundary as a map from the gauge group $\mathcal{G}$ to the boundary $\partial\mathcal{M}$. Since boundaries are almost always topologically equivalent to some $S^{n}$ (i.e. a boundary circle or sphere ), and many gauge groups are also isomorphic to some $S^{k}$, the maps $\mathcal{G}\rightarrow\partial\mathcal{M}$ often become maps $S{k}\rightarrow S^{n}$, which belong to $Hom(S^{k}, S^{n})$, and hence to the $k^{th}$ homotopy group of the manifold.  \\
The simplest examples of this are Monopole solutions \cite{Manton2004}, where the gauge group is $U(1)\simeq S^{1}$. If the manifold has a boundary which is topological $S^{1}$, then  vector potentials  $\vect{A}$ which are pure gauge on this boundary are members of $H^{1}\simeq Hom(S^{1}, S^{1}) \simeq \mathbb{Z}$, and are therefore classified (and quantised) by their winding number $k$ on the boundary circle\footnote {If there are multiple boundaries then assuming $\vect{A}$ is pure gauge on each, we can  classify it by a string of integers$ \{ k_{1}, k_{2}, ... \}$ given by its winding numbers on each separate boundary.}. This phenomenon is responsible for the quantisation of magnetic charge in the Dirac Monopole. Another important example is given by Instantons \cite{Manton2004}, where the gauge group is $SU(2)\simeq S^{3}$, and the map is from $SU(2)$ to the boundary $S^{3}$ lying in $\mathbb{R}^{4}$. These maps are also classified by $\mathbb{Z}$, and hence each is labelled by an integer known as the topological charge, or Instanton number\footnote{We might ask if we can use a gauge transformation to change this charge. Mathematically, this is possible-- we can gauge transform by some element with a non- zero charge $\kappa$,  and this will change the charge of our solution by $\pm\kappa$. Physically there is some consensus that such, `large' gauge transformations represent different solutions \cite{Landsman1997} and therefore, `gauge transformations',  in the usual sense of transformations relating physically distinct solutions, are limited to so-called, 'small gauge transformations', which do not change the boundary flux. This also means that the example we discussed above, which changes a zero-flux solution to $\vd Z=0$ to a solution with a pure gauge flux $\vect{a}$ cannot change the global flux across a boundary in a theory with topological charge. It will remain zero. }  . Physically, in both cases, this charge corresponds to the flux of $\mathcal{A}$ across the boundary, and the topological conditions force this to be quantised. This quantised charge on boundary determines the form of the solution of $\mathcal{A}$ in the interior, and therefore indirectly determines the solution for $Z$ via equation (\ref{eqn:DiracGauge}).  \\
Overall then, the use of Clifford algebras gives a clear geometric picture of the Dirac equation in terms of flux.  What is of especial interest for the Bohm-Hiley approach to quantum mechanics is the inherent nonlocality of the Dirac equation, which comes from its links to the topology of the underlying space. 
\section{The Dirac Equation and the Quantum Potential}
This section attempts to connect the Dirac equation and the Bohm approach to quantum mechanics. Bohm\cite{Bohm1952}\cite{BohmHiley1992} realised that if we write the wavefunction in polar form 
\begin{equation}\label{eqn:polarwave}
	\phi(t, \vect{x})=\rho(t, \vect{x})\ e^{\frac{i}{\hbar}S(t, \vect{x})}
\end{equation}
The Schroedinger equation 
\begin{equation}
	i\hbar\frac{\partial\psi}{\partial t}=-\hbar^{2}\nabla^{2}\psi+U\psi
\end{equation}
splits into real and imaginary parts, to eventually give \cite{BohmHiley1992}
\begin{align}\label{eqn:BohmHJ}
	\frac{\partial S}{\partial t}+\frac{(\nabla S)^{2}}{2m}+U-\frac{\hbar^{2}}{2m}\frac{\nabla^{2}\rho}{\rho}=0\\ 
	\frac{\partial \rho^{2}}{\partial t}+\nabla\cdot\Big(\rho^{2}\frac{\nabla S}{m}\Big)=0\label{eqn:BohmCont}
\end{align}
We interpret $\rho^{2}$ as the probability, and $S$ as the action, and then we can see equation (\ref{eqn:BohmCont}) is a continuity equation for the probability, and equation (\ref{eqn:BohmHJ}) as a quantum variant of the Hamilton-Jacobi equation, now with an additional term 
\begin{equation}\label{eqn:Qpot}
	Q=-\frac{\hbar^{2}}{2m}\frac{\nabla^{2}\rho}{\rho}
\end{equation}
This is the quantum potential, and the analysis of the effects of this term are at the heart of Bohm's approach to quantum mechanics\footnote{It is important to remember here that, as Basil regularly pointed out, Bohm was not a Bohmian. The motivation behind Bohmian mechanics is to explore some unexpected consequences of the standard quantum mechanical framework, as a guide to how to develop a deeper theory-- not to be a itself a final or complete description of nature. See \cite{Bohm1980} for a discussion of this point}  . The best source for a full discussion is the book written by Bohm and Basil Hiley \cite{BohmHiley1992} , but there are two features I want to bring out here. First, the quantum potential is invariant under rescalings of $\rho$, and hence of $\phi$. This means that the quantum potential does not depend on the magnitude of $\phi$, but only on its form. To point out how unusual this property is, Bohm and Hiley use the analogy of a cork in a fluid. We would be surprised should it bob up and down with the same velocity no matter how far it is from the source of a wave. We would expect the effect of the disturbance to decrease with distance, but this is precisely what does not happen with the quantum potential, indicating that it introduces an element of nonlocality into the theory.\\
The second feature is that the presence of a quantum analogue to the Hamilton-Jacobi equations allows us to define a momentum $\vect{p}=\nabla S$ in the same way as for the classical Hamilton-Jacobi equation. Bohm and Hiley claim that this describes the trajectory\footnote{For more complicated systems this must be thought of as a trajectory or probability flow line in configuration space rather than physical space. It seems as though for a single particle we can identify the two spaces} of a quantum particle.They point out that it satisfies 
\begin{equation}\label{eqn:BornSommerfeld}
	\oint \vect{p}\cdot d\vect{x}=nh
\end{equation} 
which is analogous to the Born-Sommerfeld quantisation condition; the difference being that the momentum in the Born-Sommerfeld case is defined via the classical action (using the WKB approximation), whereas here it is defined using the full quantum action. \\ 
If the Schroedinger equation is simply a non- relativistic approximation to the true dynamics, then what becomes of this analysis when we generalise to relativistic speeds? To address this question, I will first examine how the Dirac operator acts on the polar decomposition (\ref{eqn:polarwave}). Second, I will investigate the precise relation between the Dirac and Schroedinger equations, and suggest that the Schroedinger equation may still have a role to play in the relativistic case.  
\subsection{The Polar form of the Dirac Equation}
For this section, we will work in $Cl(2)$, the simplest Clifford algebra. A basis for $Cl(2)$ is
\begin{equation}
	\{ 1, e_{0}, e_{1}, e_{0}e_{1} \}
\end{equation}
where $e_{0}e_{1}=I$, the pseudoscalar, and $I^{2}=-1$. Therefore the even subalgebra $Spin(2)=\{1, I\}\cong\mathbb{C}$. This means that we can directly compare to the Schroedinger case, and make full use of tools and intuitions from complex analysis\footnote{As stated in the introduction, the same tools exist for Clifford algebras, which is why we can use them for $Cl(2)$, but it is more straightforward when they are literally identical to the complex case, rather than a higher dimensional generalisation}.  Now, let us form the wavefunction, by analogy with equation (\ref{eqn:polarwave})
\begin{equation}
	\phi(t,x)=\rho(t,x)e^{\frac{i}{\hbar}IS}
\end{equation}
where $I$ is now the pseudoscalar in $Cl(2)$. The Dirac operator in $Cl(2)$ is
\begin{equation}
	\vd=e_{0}\partial_{t}+e_{1}\partial_{x}
\end{equation}
We can calculate
\begin{equation}
	\vd\phi=\big(\vd\rho)e^{\frac{i}{\hbar}IS}+\frac{1}{\hbar}\big(\vd\cdot(IS)\big)\rho(t,x)e^{\frac{i}{\hbar}IS}=\bigg(\frac{(\vd\rho)}{\rho}+\frac{I}{\hbar}\cdot(\vd S)\bigg)\phi
\end{equation}
Now, suppose we want a holomorphic\footnote{Technically, `monogenic' since we are in $Cl(2)$ rather than $\mathbb{C}$, but I will use, `holomorphic' throughout for claritiy since $Spin(2)\cong\mathbb{C}$} wavefunction, we have 
\begin{equation}\label{eqn:DiracQPP}
	\bigg(\frac{(\vd\rho)}{\rho}+\frac{I}{\hbar}\cdot(\vd S)\bigg)\phi=0\iff \frac{(\vd\rho)}{\rho}+\frac{I}{\hbar}\cdot(\vd S)=0
\end{equation}
Provided our functions are continuous. If this equation holds then $\phi$ is holomorphic. Therefore it is always true that $\vd\phi$ is holomorphic, and the complex conjugate $\bar\phi=\rho e^{\frac{-I}{\hbar}S}$ is also holomorphic. The product of holomorphic functions is also holomorphic, and so 
\begin{equation}\nonumber
	\frac{(\vd\rho)}{\rho}+\frac{I}{\hbar}\cdot(\vd S)=\big(\vd\phi\big)\bar\phi
\end{equation}
is also holomorphic\footnote{We could also conclude this by writing it as the condition that $\vd{}.\big( ln(\rho) +\frac{IS}{\hbar}\big)=0$}. Therefore its integral around a closed curve must be zero
\begin{equation}
	\oint_{C} \bigg(\frac{(\vd\rho)}{\rho}+\frac{I}{\hbar}\cdot(\vd S)\bigg)\cdot\vect{dz}=0
\end{equation}
This gives us 
\begin{equation}
	\oint_{C}\frac{\vd\rho}{\rho}\cdot \vect{dz}=-\frac{1}{\hbar}\oint_{C}\big(\vd\cdot(IS)\big)\cdot \vect{dz}
\end{equation}
Writing the circular path $\vect{dz}$ as $\vect{z}'(t)dt$, the left hand side becomes
\begin{equation}
	\int_{t=0}^{2\pi}\frac{\vect{z}'(t)\cdot\vd\rho}{\rho}dt=\oint_{C}\frac{\rho'(z)}{\rho}dz= 2\pi I^{-1}\mathcal{N}(\rho)
\end{equation}
using the Argument Principle\footnote{ This states that 
	\begin{equation}
		\frac{1}{2\pi I}\oint_{C}\frac{\rho'(z)}{\rho}dz= \mathcal{N}(\rho)-\mathcal{P}(\rho)
	\end{equation}
	Where $\mathcal{N}(\rho)$ is the number of zeros of $\rho$ inside the contour $C$, and $\mathcal{P}(\rho)$ is the number of poles inside $C$. In this case our function is monogenic/ holomorphic within $C$, so it has no poles by definition. Hence the integral is simply equal to $\mathcal{N}(\rho)$. }
from complex analysis as a corollary to Cauchy's theorem\footnote{Which as we have said above, can be extended to Clifford algebras. Here, we are in $Cl(2)$, and so the only significant difference is that $I$ is now the pseudoscalar $e_{1}e_{2}$}. Here $\mathcal{N}(\rho)$ is the number of zeros of $\rho$ inside the contour. We can write the right hand side as
\begin{equation}
	-\frac{I}{\hbar}\oint_{C}\big(\vd S)\cdot\vect{dz}=-2\pi I\omega(\phi)
\end{equation}
By definition, where $\omega\phi$ is the winding number of $\phi$ around $C$. This must be an integer, say, $n$, and it must therefore be the case that
\begin{equation}
	\oint_{C}\big(\vd S)\cdot\vect{dz}=nh
\end{equation}
This is the same analogy to the Born-Sommerfeld condition as was found in \cite{BohmHiley1992} for the Schroedinger equation (See  equation (\ref{eqn:BornSommerfeld})). Now, putting both sides together, we have 
\begin{equation}\label{eqn:DBS}
	\mathcal{N}(\rho)=\omega(\phi)=nh
\end{equation}
In particular, adding a vector potential $\vect{A}$ to the action increases the winding number of the action by the winding number (and hence the boundary flux) of $\vect{A}$. Therefore changing the boundary flux of the vector potential by one unit changes the number of zeros of the probability $\rho$ in the interior by one as well. This is a very strong non- local effect. \\
This nonlocality, once again topological in origin, is the main result of this section. We can also comment on the two features of the Bohm equations which we highlighted above. Whilst there is no additional quantum potential term appearing, we do have the term $\frac{\vd\rho}{\rho}$ which has the same scale invariance. This term already appears in the Schroedinger case, where it is called\footnote{I am grateful to Prof. A. Sanz for bringing this to my attention} the osmotic velocity \cite{Nelson1966}\cite{BohmHiley1989}. It is linked to the study of weak measurements, as shown by Basil Hiley in  \cite{Hiley2012}. The quasi Born-Sommerfield condition (\ref{eqn:DBS}) is a new feature in the Dirac case, however. \\
Secondly, we again have the appearance of the Bohm momentum $\vd S$, though this now includes a time component, not just spatial ones as in the Schroedinger case. Nevertheless, the fact it satisfies the same Born-Sommerfeld condition implies we should give it the same interpretation. Interestingly, in the Dirac case, this is explicitly linked to the function $\frac{\vd\rho}{\rho}$ by equation (\ref{eqn:DiracQPP}), implying that the Bohm trajectories can also be calculated via $I\hbar \frac{\vd\rho}{\rho}=I\hbar\vd\big(ln(\rho)\big)$. This is a new link not present in the Schroedinger case. Future work in this area would involve using Clifford analysis to extend this analysis to the Dirac equation in higher dimensions, especially in $Cl(3,1)$. 
To conclude this section, if we consider the the full Dirac equation we get
\begin{equation}
	\vd \phi-\vect{A}\phi=m\phi
\end{equation}
Where $\vect{A}$ is the $U(1)$ vector potential. This gives
\begin{equation}
	\Big(\frac{\vd\rho}{\rho}+\frac{I}{\hbar}\cdot(\vd S)-\vect{A}-m\Big)\phi=0
\end{equation}	
This implies that 
\begin{equation}
	\frac{\vd\rho}{\rho}+\frac{I}{\hbar}\cdot(\vd\vect{S})-\vect{A}-m=0
\end{equation}
Two points arise here. First of all, integrating both sides we get
\begin{equation}
	\mathcal{N}(\rho)-\omega(\phi)-\omega(\vect{A})-\oint_{C}m dz=0
\end{equation}
as a generalisation of equation (\ref{eqn:DiracQPP}), where $\omega(\vect{A})$ is the winding number of $\mathcal{A}$. Since the winding numbers, and the number of zeroes of $\rho$ are all integers, this implies that contour integral of the mass term too, must be a multiple of $\hbar$ and hence quantised. \\
The second point comes from looking at the momentum term $\vd\phi=\frac{\vd\rho}{\rho}+\frac{I}{\hbar}\cdot(\vd S)$ . Recalling that in the classical case, we have $\vect{p}=\vd S$, we see that here the momentum picks up an extra term, the osmotic velocity $\frac{\vd\rho}{\rho}$ relative to the classical case. Comparing to the quantum Hamilton-Jacobi equation (\ref{eqn:Qpot}), this suggests that the osmotic velocity introduces a non-local quantum effect to the momentum, just as the quantum potential introduces such an effect to the energy. All this merits continued study. \\
\subsection{The relation between the Dirac and Schrödinger Equations}
The Dirac equation is often said to be the non- relativistic version of the Schroedinger equation, but this requires clarification. As Weinberg points out \cite{Weinberg1995} the Schroedinger equation describes a quantum particle, whereas the Dirac equation describes multiple particles; or, more accurately, a field with creation and annihilation operators for particle-like excitations.  \\
Additionally, the Dirac equation is the square root of a wave equation, whereas the Schroedinger equation looks like a complexified form of the heat equation
\begin{equation}\label{eqn:Heat}
	\partial_{t}\phi=\kappa\partial^{2}_{x}\phi
\end{equation}
Here, $\kappa$ is a constant measuring the rate of heat diffusion. This has even lead to the Schroedinger equation being interpreted in terms of the diffusion of probability (including by Bohm and Hiley \cite{BohmHiley1992})\footnote{They point out that the diffusive effect of the Schroedinger equation is balanced by the organising effect of the quantum potential}. \\
I suggest it is possible that, rather than the Schroedinger equation being simply a non- relativistic approximation to a Dirac equation, the two equations may have different roles. The Dirac operator describes motion in a geometric space , with the geometry determined by the underlying Clifford algebra\footnote{The original Dirac equation describes motion in Minkowski space $\mathbb{R}^{3,1}$, corresponding to the Clifford algebra $Cl(3,1)$}. The Schroedinger equation, on the other hand, is defined in configuration space\footnote{I would claim, following \cite{Hiley2011}, that we may be able to view a geometric algebra as both a geometric, physical space and a configuration space. But the point still stands} and seems to describe the dissipation of probability over time. It can also be applied to any quantum state\footnote{At least in its most general form, $\partial_{t}\phi=U\phi$, where $U$ is a time evolution operator}, whereas the Dirac equation only makes sense for describing motion in some geometry.   \\
An indication that the relationship between the two equations might be deeper than just the relativistic limit is is provided by the fact that there is a strong, topological  link between solutions of the heat equation and solutions to Dirac equations \cite{Delanghe1992}. The precise relation is determined by the Atiyah- Singer index theorem, further indicating that exploring this theorem in this context would be a useful topic for future work. \\
Another clue comes from the idea of gradient flow; mainly used by probability theorists, and gaining in interest due to its applications in machine learning \cite{Figalli2021}. Given the solutions $\phi(t, \vect{x}_{i})$ to some differential equation, and an energy functional $E(t, \vect{x}_{i})$, we define the gradient flow as
\begin{equation}
	\partial_{t}\phi=\nabla E
\end{equation}
where $\nabla$ is the usual vector gradient (in the $x_{i}$ variables). It turns out \cite{Figalli2021} that the heat equation is a gradient flow, for the energy
\begin{equation}
	E=\int \big(\nabla\phi\big)^{2} d\vect{x}
\end{equation}
Since in this case\footnote{After integration by parts} $\nabla E=\nabla^{2}\phi$, and we recover equation (\ref{eqn:Heat}).It seems to me it automatically follows that the Schroedinger equation
\begin{equation}
	i\hbar\partial_{t}\phi=-\frac{\hbar{^2}}{2m}\partial_{x}^{2}\ \phi+U\phi
\end{equation} 
is a Wick rotated\footnote{Basically we have multiplied the time variable $t$ by the imaginary unit $i$}  gradient flow equation, but for the energy functional 
\begin{equation}
	E=\int\Big[\frac{1}{2m}\big(\nabla\phi\big)^{2}+U\Big] \ d\vect{x}
\end{equation}
Now, this is a non-relativistic energy functional\footnote{The relativistic energy\cite{Taylor1966} for a single particle should be $E= \sqrt{p^{2}+m^{2}}=m\sqrt{1+p^{2}/m^{2}}$, to which the classical energy is a first order approximation in $p^{2}/ m^{2}$} but there is the matter of the Wick rotation to think about. Usually, a Wick rotation takes us from Euclidean to Minkowski space. Therefore, if the Schroedinger equation looks like a Wick rotated version of the heat equation, it is worth considering whether we can view it as a Minkowski space version of the heat Equation\footnote{There have been many attempts to derive a relativistic (or at least finite-propagation-speed) version of the heat equation- see \cite{Joseph1989}}. At first, this seems implausible, since the Schroedinger equation is a parabolic PDE and hence its solutions can have infinite speed of propagation, violating causality in Special Relativity\footnote{It is interesting to compare this to the infinite speed of information transfer implied by the quantum potential-- though I suspect that this will vanish in a future, more complete, theory}. This is easiest to to see if we consider the time independent Schroedinger equation, with Hamiltonian function $H$
\begin{equation}
	i\hbar\partial_{t}\phi=-H(\vect{x}, \vect{p})\phi\implies \ \phi=e^{\frac{i}{\hbar}H}
\end{equation}
Then (in the absence of other constraints) a non- zero solution at $t=0$ implies non- zero solutions for any other $(t, \vect{x})$. Now, this can be avoided by choosing a relativistic Hamiltonian\footnote{For example, if we choose as the Dirac Hamiltonian $\gamma_{0} mc^{2}+\sum_{i}^{3}\gamma_{i}p_{i}$ the the Schroedinger equation becomes the Dirac equation \cite{Dirac1988}} which for fixed $t$ is only non- zero for suitably small $\vect{x}$ in the forward light cone. It is clear that whilst the Schroedinger equation does not necessarily give relativistic solutions., it does not exclude them either. I will now give a more concrete example of a relativistic (indeed, lightlike) solution to Schroedinger's equations. \\
Given that we are looking for a Minkowski space solution, let's start with an wave ansatz
\begin{equation}\label{eqn:ansatz}
	\phi(t, \vect{x})=e^{\frac{i}{\hbar}\big(p_{0}ct\pm\vect{p}\cdot\vect{x}\big)}\equiv e^{\frac{i}{\hbar}\big(E_{0}t\pm\vect{p}\cdot\vect{x}\big)}
\end{equation}
where $\vect{x}$ is a position vector in $\mathbb{R}^{3}$. Assume that our wave is a solution\footnote{Note that the sign $\pm$ in the ansatz corresponds to the usual positive and negative energy Klein-Gordon solutions} to the wave equation
\begin{equation}\label{eqn:wave}
	\partial_{t}^{2}\phi-\sum_{i=1}^{3}\partial_{i}^{2}\phi=0
\end{equation}
and so satisfies $E_{0}^{2}-\lvert \vect{p}\rvert^{2}=0$. We can now plug this solution in to Schroedinger's equation. Let us assume there is no potential, so $U=0$, and let us make some modifications to the constants. This will not change the dynamics but will help with the interpretation of the equation.  The mass term $\frac{1}{2m}$ is definitely not relativistic, so I will replace it with $E_{0}$, since this component is the boost of the rest mass. Then our equation becomes
\begin{equation}
	i\hbar\partial_{t}\phi=-\frac{\hbar{^2}}{E_{0}}\partial_{x}^{2}\phi
\end{equation} 
Plugging in our ansatz (\ref{eqn:ansatz}), we get 
\begin{equation}
	E_{0}=\frac{\lvert\vect{p}\rvert^{2}}{E_{0}}\implies E_{0}^{2}-\lvert \vect{p}\rvert^{2}=0
\end{equation}
which follows from the fact that our ansatz solves the wave equation (\ref{eqn:wave}). A general solution will then be a linear combination of such waves. \\
We might worry here that the Schroedinger equation didn't contribute anything. However, the factor $\frac{1}{\hbar}$ in the ansatz (\ref{eqn:ansatz}) is not motivated by the wave equation-- it comes from the need to cancel the $\hbar$'s in the Schroedinger equation. It also has a nice physical consequence.  Taking the positive energy case, and comparing our wave function $e^{\frac{i}{\hbar}\big(E_{0}t-\vect{p}\cdot\vect{x}\big)}$ to the standard form $e^{i(\omega t-\vect{k}\cdot\vect{x})}$, where $\omega$ is the angular frequency and $\vect{k}$ gives the wavenumber, we have
\begin{align}
	E_{0}=\hbar\omega\\ 
	p_{i}=\hbar k_{i}
\end{align}
which are precisely the de Broglie equations.  So if we modify the coefficients of the Schroedinger equations, we can describe a lightlike de Broglie wave. Now, for a massive particle, we start with an ansatz which is a solution to the Klein- Gordon equation
\begin{equation}\label{eqn:KG}
	\partial_{t}^{2}\phi-\sum_{i=1}^{3}\partial_{i}^{2}\phi=m^{2}\hbar^{2}\equiv M^{2}
\end{equation}
This will be equation (\ref{eqn:ansatz}), but now with $E_{0}^{2}-\lvert\vect{p}\rvert^{2}=M^{2}$. Then to make this work with Schroedinger's equation, we must add a potential $U=M^{2}/E_{0}$. Then our Schroedinger equation is
\begin{equation}\label{eqn:Schmod}
	i\hbar\partial_{t}\phi=-\frac{\hbar{^2}\partial_{x}^{2}\phi + M^{2}}{E_{0}}
\end{equation}
Again, plugging in the ansatz gives
\begin{equation}
	E_{0}=\frac{\lvert\vect{p}\rvert^{2}+M^{2}}{E_{0}}\implies E_{0}^{2}-\lvert \vect{p}\rvert^{2}=M^{2}
\end{equation}
Which is true since our ansatz this time is a solution to the massive Klein- Gordon equation (\ref{eqn:KG}). We have the same De Broglie relations as for the massless case. In the case of the heat equation (\ref{eqn:Heat}), the coefficient of $\nabla^{2}$, $\kappa$, gave the rate of heat dissipation. By analogy, here the rate of probability dissipation is inversely proportional to $E_{0}$. This makes sense, since larger, and hence more classical, systems are more coherent and less prone to a spreading of the wavefunction. \\
Now, these solutions are somewhat ad hoc \footnote{though I suspect they can be made less ad hoc and more systematic with further investigation},  however I think it is striking that with some minor modifications to the coefficients in the Schroedinger equation\footnote{If we had not modified the coefficients we would have got the wave (restricting to 2d for simplicity), $\phi=e^{\frac{i}{\hbar}(t\pm2mx)}$ satisfying $\partial^{2}_{t}\phi-\partial^{2}_{x}\phi=\hbar^{2}(1-4m^{2})$, which is still a hyperbolic (i.e. finite propagation speed) solution, but does not seem particularly physical }, we can describe both massive and lightlike De Broglie waves. The Schroedinger equation allows for relativistic (and even lightlike) solutions, and therefore its obsolescence in the relativistic case  cannot be automatically assumed, especially with the underlying links between the heat equation and Dirac equation\footnote{As a final point, we can see how the Bohm form of the Schroedinger equation is affected by the modified coefficients in equation (\ref{eqn:Schmod}). It is easy to see that we get
\begin{align}
	\frac{\partial S}{\partial t}+\frac{(\nabla S)^{2}+M^{2}}{E_{0}}-\frac{\hbar^{2}}{E_{0}}\frac{\nabla^{2}\rho}{\rho}=0\\ 
	\frac{\partial \rho^{2}}{\partial t}+\nabla\cdot\Big(\rho^{2}\frac{2\nabla S}{E_{0}}\Big)=0
\end{align}
Whilst I do not think there is anything immediately illuminating here, I include them for completeness.}.
\section{Conclusion}
So, after this survey, what have we learnt? I would say that the main points are the following:
\begin{enumerate}
	\item The Dirac operator in the geometric algebra formalism can be applied to any geometric object, not just spinor fields
	\item The Dirac equation can be interpreted in terms of flux, with $\vd\cdot Z$ describing flux internal to $Z$, and $\vd\wedge Z$ describing external flux. 
	\item Solutions to the Dirac equation are often nonlocal; either due to Hodge theory in Euclidian space, or due to Soliton- like effects from gauge theories. Further work is needed here- both to integrate the Atiyah- Singer index theorem into geometric algebra, and to look the extension of Hodge theory (or at least its links between the kernel of the Dirac equation and homology) to Minkowski space.-- but  the Dirac equation seems to be fundamentally nonlocal and holistic, as the Bohm/Hiley approach would predict.
	\item Using the polar decomposition of the wavefunction with the Dirac equation yields physically interesting results. Once more there is nonlocality with its roots in topology. 
	\item The relationship between the Dirac and Schroedinger equations is more complicated than simply saying the Dirac equation is relativistic, and the Schroedinger equation isn't. Again, more work is required to fully understand this-- especially around the link that the Atiyah- Singer index theorem sets up between the Dirac and heat equations. 
\end{enumerate}
Returning to our list of key features of the Bohm- Hiley approach, the analysis of the  Dirac equation in this paper indicates new approaches to the Schroedinger equation, and the continued usefulness of the polar decomposition of the wavefunction in the Dirac context. It offers new approaches to nonlocality in quantum mechanics arising from topological effects\footnote{There is of course a whole research programme in Topological Quantum \cite{Simon2023}}. Naturally, Basil's interests also extended here, especially to the role of Braid groups, and the work of Lou Kauffman \cite{Kauffman1994}. I am not aware of a large role for the the Dirac equation and its topological properties in the existing  topological quantum literature, but I am sure that such links exist and will prove highly fruitful\footnote{The fact that Dirac operators quantise the well studied Chern-Simons theory \cite{Delanghe1992},  is an obvious place to start}. \\
The corroborations of the Bohm-Hiley approach to quantum mechanics from this brief study of Dirac operators both offer additional support for that approach, and, since Clifford algebras are intrinsically tied to Dirac operators,  indicate the validity of Basil's insight that Clifford algebras are a powerful language for developing it further. Basil's work and contribution to physics during his lifetime were immense and inspiring, but I feel that his legacy will only grow in the future as the projects he began are continued and  the vision of physics he devoted his life to becomes clearer. 
\bibliographystyle{siam}
\bibliography{mathspapers2025}

\end{document}